\documentclass[twocolumn,showpacs,preprintnumbers,aps,pre,amsmath,amssymb]{revtex4-1}

\usepackage{graphicx}
\usepackage{dcolumn}
\usepackage{bm}
\usepackage{multirow}

\usepackage{amsmath,amssymb,amsthm}
\providecommand{\abs}[1]{\lvert#1\rvert}
\providecommand{\pdiff}[1]{\frac{\partial}{\partial #1}}
\providecommand{\pdiffarg}[2]{\frac{\partial #1}{\partial #2}}

\providecommand{\V}[1]{\boldsymbol #1}
\providecommand{\T}[1]{\overleftrightarrow{\boldsymbol #1}}
\providecommand{\Av}[1]{\left\langle #1 \right\rangle}
\providecommand{\C}[1]{{\cal #1}}
\providecommand{\tx}[1]{\text{#1}}

\providecommand{\kBT}[0]{k_\tx{B} T}

\providecommand{\dd}{\text{d}}
\providecommand{\hho}{{\text{H}_2\text{O}}}

\providecommand{\AdressTUM}{Physik Department, Technische Universit\"at M\"unchen, 85748 Garching, Germany}
\providecommand{\AdressFUB}{Fachbereich Physik, Freie Universit\"at Berlin, 14195 Berlin, Germany}
\providecommand{\AdressIPST}{Institute for Physical Science and Technology, University of Maryland, College Park, MD 20742, USA}

\begin{document}

\title{Friction contribution to water-bond breakage kinetics}
\author{Yann von Hansen}%
\author{Felix Sedlmeier}%
\author{Michael Hinczewski}%
\altaffiliation[Now at: ]{\AdressIPST}%
\affiliation{\AdressTUM}%
\author{Roland R. Netz}%
\email{email: rnetz@physik.fu-berlin.de}%
\affiliation{\AdressFUB}%

\date{\today}
\begin{abstract}
Based on the trajectories of the separation between  water molecule pairs from  MD simulations,
we investigate the bond breakage dynamics in bulk water. From the spectrum of mean first-passage times,
the Fokker-Planck equation allows us to derive the diffusivity profile along the separation coordinate and thus
to unambiguously  disentangle the effects of  free-energy and local friction on the separation kinetics.
For  tightly coordinated water the friction is six times higher than in bulk, which can be interpreted in terms
of  a dominant reaction path that involves additional orthogonal coordinates.
\end{abstract}


\maketitle

\section{Introduction}
\label{sec:Introduction}
The unique properties of liquid water are relevant for a broad range of processes in biology, chemistry, and physics, 
as well as for technological applications~\cite{Ball2008}. 
A prominent goal of recent research has been to relate macroscopic properties (among those the notable anomalies and singularities)
to the microscopic structure and thus to the hydrogen (H) bonding pattern between individual water molecules~\cite{Stillinger1980}.
This goal has only partly been achieved.
Indeed, even for the most elementary kinetic process of breaking a single H-bond between two water molecules
that are embedded in the bulk liquid matrix, various viewpoints exist:
In an early application of transition path sampling, it was found that in roughly half of the cases of an H-bond breaking event a new bond forms
right afterwards~\cite{Csajka1998}, supporting Stillinger's switching-of-allegiance description of the local water dynamics~\cite{Stillinger1980}. 
In later simulation works, the water reorientation during  this H-bond switching was shown to occur quite abruptly~\cite{Laage2008},
in line with the  pronounced rotational-translational motion coupling of individual water molecules~\cite{Svishchev1993}.
The non-exponential H-bond relaxation was shown to be due to a coupling of bond making/breaking dynamics
and the relative diffusion of water pairs~\cite{Luzar1996}, but not related to the local environment of H-bond forming 
water molecules~\cite{Luzar1996b}, which is surprising in light of the above mentioned H-bond switching scenario.
Clearly, the H-bond dynamics is intimately related to the kinetics of e.g. protein folding~\cite{Rhee2004} 
or solute dissociation~\cite{Geissler1999}, so 
clarifying the kinetics of the binding and unbinding of water molecules is without doubt of fundamental importance.

The concept of diffusion along a reaction coordinate (RC) has been fruitful for gaining insight into the underlying mechanisms of high-dimensional dynamics as in the case of protein folding, for which various approaches to identify suitable RCs~\cite{Du1998,*Rhee2005,E2010} and to locate or characterize  transition states~\cite{Bolhuis2002,Best2005,Noe2009} have been developed.
Here, we choose the separation between two water molecules as the naive RC and show that a consistent description of the dynamics along the separation coordinate can be obtained.
In fact, our stochastic analysis in terms of the Fokker-Planck (FP) equation with coordinate-dependent free-energy and  diffusivity allows us to quantify to which extent degrees of freedom that are orthogonal to our chosen RC are involved in  the reaction.
As a main result, we find the relative translational friction in the first coordination shell to be more than six-fold  increased compared to bulk water.
Application of transition rate theory without taking this local friction change into account underestimates typical bond breakage times by a factor of two.
\begin{figure}
  \includegraphics[width=\columnwidth]{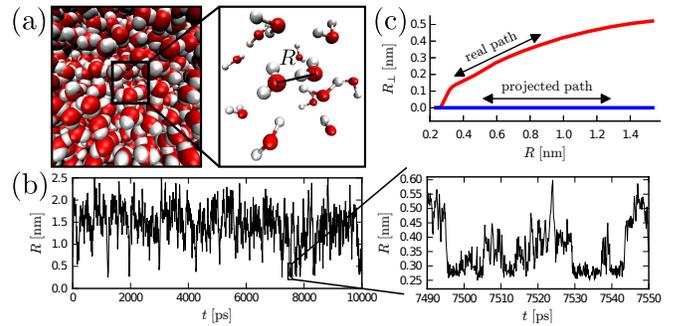}
\caption{\label{fig:TimeSeries}[Color online] (a)~Simulation snapshot visualized using VMD~\cite{Humphrey1996}: 
the coordinate $R$ in the enlarged section is defined as the radial separation between the oxygen atoms.
(b)~Typical time series of $R$, the magnification reveals fluctuations on the subpicosecond scale. 
A simulation \textit{movie} is provided as supplementary material~\cite{Suppl}.
(c)~Illustrative typical reaction path involving in addition to the separation $R$ an orthogonal component 
$R_\perp$ (cf. text in Sec.~\ref{sec:Interpretation_Diffusivities}).}
\end{figure}
Our analysis is based on $10~\tx{ns}$ long trajectories of the separation $R$ between the oxygen atoms of water pairs
provided by molecular dynamics (MD) simulations of the standard three point charge water model SPC/E~\cite{Berendsen1987};
see  Fig.~\ref{fig:TimeSeries}a) and b) for a snapshot and an example trajectory; a simulation movie is provided as supplementary material~\cite{Suppl}.
We do not check for the presence of H-bonds --- which would introduce an element of arbitrariness due to the H-bond definition~\cite{Luzar2000,*Kumar2007} --- but rather base our discussion solely on the separation $R$; strictly speaking we do 
not consider H-bond kinetics but more generally water-bond kinetics.

The paper is organized as follows: We start by reviewing the FP equation in Sec.~\ref{sec:FP_radial_dynamics}, on which our analysis is based. The simulation setup is described in Sec.~\ref{sec:MD_simulations}, details regarding the trajectory analysis are given in Sec.~\ref{sec:MFPTs_from_MD}. Results are presented in Sec.~\ref{sec:Results} and discussed in Sec.~\ref{sec:Discussion}. A summary of our main findings is given in Sec.~\ref{sec:Conclusions}, while technical aspects are covered in the appendices.

\section{Fokker-Planck Equation for Radial Dynamics}
\label{sec:FP_radial_dynamics}
In the overdamped limit, the FP equation in three dimensions describes the time evolution of the probability density $\Psi$ of observing a vectorial separation $\V r$ at time $t$,
 \begin{equation}
\label{eq:FP3Da}
\pdiff{t}\Psi(\V r,t)=-\nabla\cdot\V J(\V r),
\end{equation}
where the probability flux density
\begin{equation*}
\V J(\V r)=-\Psi(\V r,t)\T \mu_{3\tx{D}} (\V r)\cdot\nabla U(\V r)-\T D_{3\tx{D}}(\V r)\cdot\nabla\Psi(\V r,t),
\end{equation*}
has two contributions: (i)~the overdamped motion due to an (effective) potential $U$ and (ii)~diffusion with a (possibly) position dependent diffusivity tensor $\T D_{3\tx{D}}$.
Using the Einstein relation $\T D_{3\tx{D}}=\kBT \T \mu_{3\tx{D}}$ connecting mobility and diffusivity, Eq.~\ref{eq:FP3Da} can be rewritten as
\begin{equation}
\label{eq:FP3Db}
 \pdiff{t}\Psi(\V r,t)=\nabla\cdot\left(\tx{e}^{-\beta U(\V r)}\T D_{3\tx{D}} (\V r)\cdot\nabla \left(\tx{e}^{\beta U(\V r)}\Psi(\V r,t)\right)\right),
\end{equation}
where $\beta\equiv1/(\kBT)$ denotes the inverse thermal energy. Within this paper, we concentrate on the relative dynamics of two water molecules along their radial distance $R$. Since the diffusion tensor
\begin{equation}
 \T D_{3\tx{D}}=\begin{pmatrix} D & 0 & 0 \\ 0 & D_\Theta & 0 \\ 0 & 0 & D_\Phi \end{pmatrix},
\end{equation}
remains diagonal when introducing spherical coordinates $(R,\Theta,\Phi)$ and the effective inter-molecular potential $U$ due to symmetry depends on $R$ only, the angular coordinates can be integrated out~\cite{Agmon1991}. The time evolution of the radial probability distribution
\begin{equation}
 P(R,t)\equiv\int_{0}^{2\pi}\dd\Phi\int_{0}^{\pi}\dd\Theta\;\sin\Theta R^2 \Psi(R,\Theta,\Phi,t),
\end{equation}
specifying the probability of finding a radial distance $R$ at time $t$, is described by the simpler equation
\begin{equation}
\label{eq:FP1Drad}
 \pdiff{t} P(R,t)=\pdiff{R}\left(R^2\tx{e}^{-\beta U(R)}D(R)\pdiff{R}\left(\tx{e}^{\beta U(R)}\frac{P(R)}{R^2}\right)\right),
\end{equation}
where the pair radial diffusivity $D$ may depend on $R$.
It is useful to absorb the factors $R^2$ in Eq.~\ref{eq:FP1Drad} by defining a free-energy $F\equiv U-2\kBT\log R$~\cite{Agmon1991} to recover the usual form of the one-dimensional FP equation~\cite{Risken,Gardiner}
\begin{equation}
\label{eq:Fokker-Planck}
\begin{split}
\frac{\partial}{\partial t} P(R,t) 
=\frac{\partial}{\partial R} \left(D(R) \tx{e}^{-\beta F(R)}  \frac{\partial}{\partial R}  \left(P(R,t) \tx{e}^{\beta F(R)}\right)\right). 
\end{split}
\end{equation}
The free-energy $F(R)=-\kBT\log{\Av{P(R)}}$ is obtained by Boltzmann inversion of the equilibrium probability $\Av{P(R)}$.
Determining  $D(R)$ is more subtle: Different procedures have been proposed in the context of protein folding~\cite{Pogorelov2004,*Yang2007,Best2010,Hinczewski2010JCP,vonHansen2010JPCB} or interfacial water diffusion~\cite{Liu2004,*Sega2005,*Wick2005,Mittal2008,Sedlmeier2011JSP}.
Here, we obtain $D(R)$ directly from measured  mean first-passage times (MFPTs); for diffusive dynamics described by Eq.~\ref{eq:Fokker-Planck}, the MFPT $\tau_\tx{fp}$ of first reaching a separation $R_\tx{t}$ when starting off from $R$ is given by~\cite{Weiss1966}
\begin{equation}
\label{eq:MFPT}
\tau_\tx{fp}(R, R_\tx{t}) = \int^{R_\text{t}}_{R} \dd R^\prime\,\frac{\tx{e}^{\beta F(R^\prime)}}{D(R^\prime)} 
\int^{R^\prime}_{R_\text{min}} \dd R^{\prime\prime} \tx{e}^{-\beta F(R^{\prime\prime})},
\end{equation}
assuming a reflective (zero-flux) boundary condition at $R_\tx{min}<R<R_\tx{t}$.
By differentiation, one readily gets~\cite{Hinczewski2010JCP}
\begin{equation}
\label{eq:MFPTmethod}
D(R)=-\frac{\tx{e}^{\beta F(R)}}{\partial \tau_\tx{fp}(R, R_\tx{t}) / \partial R}
\int^{R}_{R_\text{min}} \dd R^{\prime} \tx{e}^{-\beta F(R^{\prime})}.
\end{equation}
Extracting MFPT curves $\tau_\tx{fp}$ from simulation data thus allows to determine the separation dependent diffusivity $D(R)$ governing the dynamics in the free-energy landscape $F(R)$; resulting diffusivity profiles are presented in Sec.~\ref{sec:Results}.

\section{Methods}
\subsection{Simulation Setup}
\label{sec:MD_simulations}
MD simulations of the SPC/E~\cite{Berendsen1987} water model are performed with the Gromacs simulation package~\cite{Hess2008,*vanderSpoel2005}. Systems consisting of 895 and 2180 water molecules are simulated in a cubic box with periodic boundary conditions. At $T=300~\tx{K}$ this corresponds to box sizes of roughly $3.0\times3.0\times3.0~\mathrm{nm}^3$ and $4.0\times4.0\times4.0~\mathrm{nm}^3$. Most simulations are performed using 895 molecules; results for target separations $R_\tx{t}=1.9~\tx{nm}$ stem from simulations involving 2180 molecules. We perform simulations at temperatures of $T = 280, 300, 320$ and $340~\tx{K}$ for the small system and at $T = 300$~K for the large system at a pressure of $P = 1~\tx{bar}$. At each temperature the system is equilibrated first in an $NVT$ ensemble (constant particle number, volume and temperature) for $100~\tx{ps}$ and then in an $NPT$ ensemble (constant particle number, pressure and temperature) for $100~\tx{ps}$. Production runs are performed subsequently for $t = 10~\tx{ns}$ and configurations are saved each $10$~fs for the small system and each $100$~fs for the large system. A Berendsen weak coupling thermostat and barostat~\cite{Berendsen1984} with a relaxation time of $\tau = 1.0~\tx{ps}$ is used for temperature and pressure control. All non bonded interactions are cut off at a radius of $R_{\tx{c}} = 0.9\,\tx{nm}$. Long-range electrostatic interactions are treated by the particle mesh Ewald method~\cite{Essmann1995,*Darden1993} with tinfoil boundary conditions. An analytic long-range correction for the Lennard-Jones interaction is applied to energy and pressure.

The simulation movie available as supplementary material~\cite{Suppl} visualizes the relative dynamics of a chosen water pair in the MD simulation ($T=300~\tx{K}$) and was created using VMD~\cite{Humphrey1996}.

\subsection{Molecular Dynamics Data Analysis}
\label{sec:MFPTs_from_MD}
The diffusion constant $D_\hho$ of a single water molecule is  determined from the long-time limit of the single-molecule mean squared displacement (MSD), $D_\hho=\lim_{t\to\infty}\Av{(\V r(t)-\V r(0))^2}/(6t)$, as detailed in App.~\ref{sec:H2O_Diffusion_Coefficients}.

The relative dynamics of all pairs of water molecules within the $10~\tx{ns}$ long MD trajectory are resolved using their minimal image distance with a spatial resolution of $\Delta R=0.002~\tx{nm}$ and a temporal resolution of $\delta t=20~\tx{fs}$.
All paths starting within a distance $\Delta R/2$ from $R$ and crossing $R_\tx{t}-\Delta R/2$ at a time $t_\tx{fp}$ later contribute to the mean first-passage time (MFPT) $\tau_\tx{fp}=\Av{t_\tx{fp}}$ and to the first-passage time (FPT) distribution $f_\tx{fp}$.
Due to the periodicity of the system, the relative dynamics along the coordinate $R$ is only meaningful for $R\leq L/2$ with box size $L$; we therefore only consider target distances $R_\tx{t}<L/2$.
Note that the absolute values of the MFPT curves $\tau_\tx{fp}(R,R_\tx{t})$ sensibly depend on the time resolution $\delta t$ of the underlying trajectory; this sensitivity is discussed in App.~\ref{sec:MFPTs_time_resolution_model}.

The derivative $\partial\tau_\tx{fp}(R,R_\tx{t})/\partial R$ in Eq.~\ref{eq:MFPTmethod} is determined by fitting a straight line to $\tau_\tx{fp}$ within a region of width $0.032~\tx{nm}$ around $R$ (corresponding to 17 data points). The width of the region was empirically found to smooth out the statistical noise in the MFPT curves without hiding the relevant variations of the diffusivity.
The integral in Eq.~\ref{eq:MFPTmethod} is evaluated numerically, the equilibrium distribution $\Av{P(R)}$ is linearly interpolated and the reflective boundary set to $R_\tx{min}=0.235~\tx{nm}$.
Applying the same kind of procedure based on simulations of 2180 molecules in a cubic box of edge length $L\approx4~\tx{nm}$ allows to consider targets $R_\tx{t}$ up to $1.9~\tx{nm}$ without introducing artifacts due to the periodicity of the simulation box and thus resolving the diffusivity $D$ over a larger range of separations $R$; finite size effects in the diffusivity profiles were not observed.

 We thoroughly checked that for all numerical steps of the data analysis, varying the spatial and temporal resolutions as well as the position of the reflective boundary $R_\tx{min}$ had no significant impact on the resulting diffusivity profiles.

\section{Results}
\label{sec:Results}
\subsection{Mean First-Passage Times and Diffusivity Profiles}
\begin{figure}
 \includegraphics[width=.9 \columnwidth]{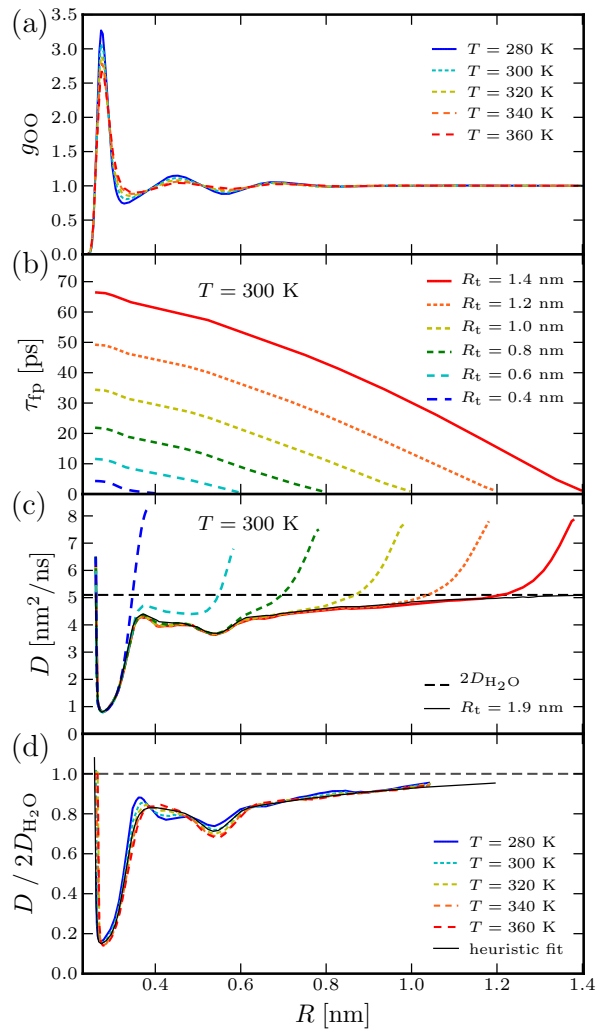}
\caption{\label{fig:rel_diff}[Color online] (a)~Pair correlation function $g_\tx{OO}$   from MD simulations for various temperatures.
(b)~MFPT curves $\tau_\tx{fp}$ from MD data for $T=300~\tx{K}$ and 
several target separations  $R_\tx{t}$. (c)~Diffusivity profiles  $D$ at $300~\tx{K}$ from the distribution in (a), the MFPTs in (b) and Eqs.~\ref{eq:MFPTmethod} and \ref{eq:rel_F_g_rad} (same colors as in (b)).
(d)~Diffusivity profiles rescaled by the bulk diffusivity $2~D_\hho$ for various temperatures.}
\end{figure}

Fig.~\ref{fig:rel_diff}a) shows the pair-correlation function $g_\tx{OO}(R)$ for different temperatures, the maxima indicating the positions of the respective coordination shells.
The free-energy
\begin{equation}
\label{eq:rel_F_g_rad}
F(R)=-2\kBT\log{R}-\kBT\log{\left(g_\tx{OO}(R)\right)},
\end{equation}
exhibits a barrier of about $1 \kBT$ for crossing from the first to the second coordination shell as seen in Fig.~\ref{fig:react}a).
MFPT curves $\tau_\tx{fp}$ extracted from the simulation data for targets $R_\tx{t}$ ranging from $0.4$ to $1.4~\tx{nm}$ 
for $T=300~\tx{K}$ are shown in Fig.~\ref{fig:rel_diff}b). They are converted, using Eq.~\ref{eq:MFPTmethod}, into  diffusivity profiles $ D(R)$ shown 
in Fig.~\ref{fig:rel_diff}c); details of the trajectory analysis are given in Sec.~\ref{sec:MFPTs_from_MD}.
There is rather good agreement between the curves for different target separations $R_t$, which is strictly expected only for 
a pure Markovian process as described by a one-dimensional FP equation. As will be discussed later on, this suggests that
 water bond breakage, defined as the passage from the first to the second coordination shell,
is to a good approximation Markovian. The deviations seen when $R \rightarrow R_t$ 
are expected, since on the short spatial scales associated with those first-passage events water motion is not diffusive; in fact, the crossover between ballistic and diffusive motion of single water molecules occurs at length scales of around $0.1~\tx{nm}$ (cf. Fig.~\ref{fig:MSD_single} in App.~\ref{sec:H2O_Diffusion_Coefficients}).
For increasing separation all curves saturate at a value equal to twice the diffusion constant of a single water
molecule, $D(R \rightarrow \infty) = 2 D_\hho \approx5.1~\tx{nm}^2/\tx{ns}$ (denoted by a broken line), as expected.

As our main finding, the diffusivity profile exhibits a pronounced drop within the first coordination shell and
reaches a minimum value of $D \approx0.79~\tx{nm}^2/\tx{ns}$ about six times smaller than in bulk, while factors $\sim2$ were previously observed in simpler systems~\cite{Emeis1970,*Fehder1971}.
The thin, black line in Fig.~\ref{fig:rel_diff}c) was obtained by evaluating MFPTs to a target separation $R_\tx{t}=1.9~\tx{nm}$ based on simulation data of the larger box with edge length $L\approx4.0~\tx{nm}$.

Diffusivity profiles corresponding to distinct target separations $R_\tx{t}$ deviate from each other in two respects: (i) non-Markovian dynamics on short time and length scales lead to modifications for $\abs{R-R_\tx{t}}\lesssim0.25~\tx{nm}$ as discussed above, and (ii) the statistical uncertainty increases with increasing $\abs{R-R_\tx{t}}$ due to a decreasing number of recorded transition events contributing to the corresponding MFPTs.
Smooth and reliable diffusivity profiles are thus obtained by joining the regions $R_\tx{t}-0.45~\tx{nm}\leq R<R_\tx{t}-0.35~\tx{nm}$ of the diffusivity profiles corresponding to targets $R_\tx{t}=0.6,0.7,\dots,1.4~\tx{nm}$.
The resulting diffusivity profiles rescaled by twice the bulk diffusion constant $D_\hho$ are shown in Fig.~\ref{fig:rel_diff}d) for various temperatures; the values of $D_\hho$ at different temperatures are in good agreement with previous simulation estimates and experiments as seen in Table~\ref{tab:diff-constants}.
\begin{table}
\caption{%
\label{tab:diff-constants}Temperature dependence of the diffusion coefficient $D_\hho$ of a single water molecule in bulk water. Simulation results for the SPC/E water model obtained by evaluation of the long time MSD (cf.~App.~\ref{sec:H2O_Diffusion_Coefficients}) are compared to results from previous simulation studies and to experimental findings (both with references).}
\begin{tabular*}{0.5\textwidth}{@{\extracolsep{\fill}} c | c c}
\hline\hline
\multirow{2}{*}{$T\;[\tx{K}]$}& \multicolumn{2}{c}{$D_\hho\;[\tx{nm}^2/\tx{ns}]$}\\ 
&Simulations (SPC/E) & Experiments\\
\hline
278& --- & $1.313$~\cite{Mills1973}\\
280 & $1.60~\pm~0.02$  & 1.44~\cite{Price1999}\\
298 & 2.75~\cite{Mark2001}, 2.70~\cite{Sendner2009}&$2.22-2.61$~\cite{Eisenberg,Mills1973,Price1999}\\
300 & $2.55~\pm~0.05$  & ---\\
318 & --- &3.575~\cite{Mills1973}\\
320 & $3.70~\pm~0.05$  & ---\\
340 & $5.08~\pm~0.05$  & ---\\
360 & $6.60~\pm~0.05$  & ---\\
\hline\hline
\end{tabular*}
\end{table}

Interestingly, deviations over a temperature span of $80~\tx{K}$ are very small; the main features of the profile, including the six-fold decrease within the first coordination shell, are accurately described by the heuristic formula
\begin{equation}
\label{eq:D_rel_ref}
\begin{split}
D(R)&\approx2D_\hho\biggr(10.76-0.68\, \tx{e}^{-9 \tilde R/4}-0.1\,\tx{e}^{-\frac{1}{5} (27-50 \tilde R)^2}\\
&+10 \tanh \left(50(1-4 \tilde R)\right)-0.34 \tanh\left(13.2-40 \tilde R\right)\\
&+0.1 \tanh\left(4.1-10 \tilde R\right)\biggr),
\end{split}
\end{equation}
where $\tilde R\equiv R/\tx{nm}$; Eq.~\ref{eq:D_rel_ref} is shown as thin black line in Fig.~\ref{fig:rel_diff}d).
From  the Arrhenius-like temperature dependence of the bulk diffusion coefficient
(cf. Fig.~\ref{fig:Diff_single} in App.~\ref{sec:H2O_Diffusion_Coefficients}) it follows  that the entire diffusivity profile obeys an  Arrhenius law.

\subsection{Maxima in the MFPT-Curves at Small Separations}
\label{sec:nonmonotonous_MFPTs}
\begin{figure*}
\includegraphics[width=.9\columnwidth]{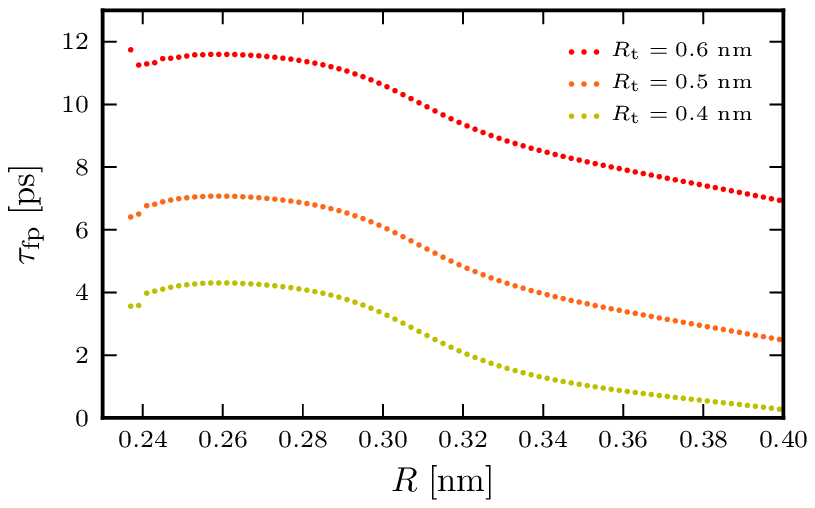}
\hspace{2em}
\includegraphics[width=.9\columnwidth]{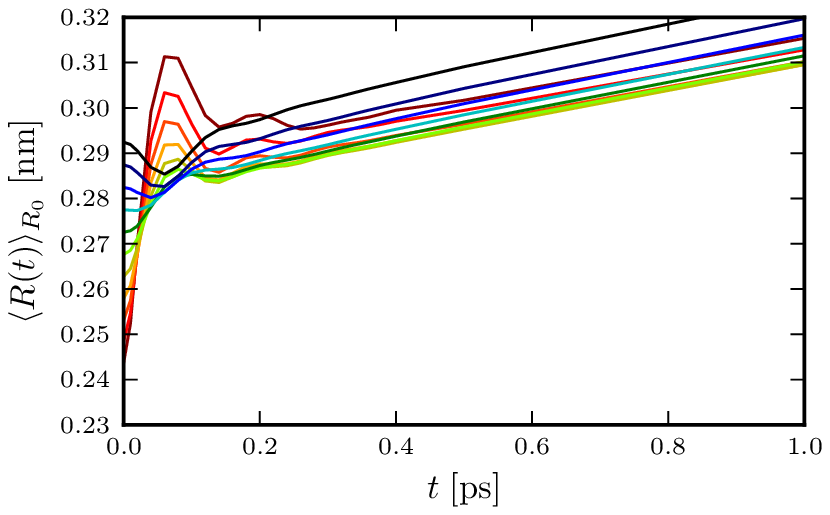}
\caption{\label{fig:Nonmonotonous_MFPTs}[Color online] \textit{Left}:~Enlarged view of the MFPT curves $\tau_\tx{fp}$ from MD simulation data at small separations (same data as in Fig.~\ref{fig:rel_diff}b)). \textit{Right}:~Average oxygen-oxygen separations $\Av{R(t)}_{R_0}$ for water pairs with defined initial separation $R_0$ at time $t=0$.
}
\end{figure*}

According to Eq.~\ref{eq:MFPT} the MFPT-curve $\tau_\tx{fp}(R,R_\tx{t})$ is a strictly decreasing function of $R$; in contrast, as can be seen in the left panel of Fig.~\ref{fig:Nonmonotonous_MFPTs}, which shows a close-up of the MFPTs of Fig.~\ref{fig:rel_diff}b) at small separations, the MFPT curves obtained from MD simulation data show a maximum at separations $R\approx0.26~\tx{nm}$.
Since according to Eq.~\ref{eq:MFPTmethod} a vanishing / positive slope of a MFPT-curve implies a diverging / negative diffusivity, the concept of Markovian dynamics obviously breaks down at such small separations. The diffusivity profiles in Fig.~\ref{fig:rel_diff}c) and d) are therefore only resolved for separations $R\geq 0.265~\tx{nm}$.

Though being counterintuitive at first sight, these maxima in the MFPTs can easily be understood by considering the average oxygen-oxygen separation $\Av{R(t)}_{R_0}$ of an ensemble of water pairs starting with defined initial separation $R_0$ at time $t=0$. 
SPC/E-water molecules interact via Coulomb and via Lennard-Jones (LJ) interactions: for small separations the repulsive part of the LJ-potential significantly contributes to the total energy of a water pair (for $R=0.25~\tx{nm}$ and $T=300~\tx{K}$:  $U_\tx{LJ}\approx13.5~\kBT$). The corresponding water pair is thus expected to be quickly driven apart due to the repulsive LJ-force. The right panel of Fig.~\ref{fig:Nonmonotonous_MFPTs} indeed reveals that the average distance between water molecules starting at separations $R_0\lesssim0.25~\tx{nm}$ increases strongly within fractions of picoseconds. The oscillations seen in the trace for $R_0\approx0.245~\tx{nm}$ nicely match the time scale of inter-oxygen vibrations, which was found to be on the order of $0.1-0.2~\tx{ps}$~\cite{Luzar1996b}. Due to the repulsive LJ-interaction for $R\lesssim0.25~\tx{nm}$, the average separations of water pairs starting out at $R_0\approx0.245~\tx{nm}$ and at $R_0\approx0.282~\tx{nm}$ are very similar on time scales $t\gtrsim1~\tx{ps}$ and exceed the average separation of pairs starting in the range $0.255~\tx{nm}<R_0<0.275~\tx{nm}$; the maxima observed in the MFPT curves of Fig.~\ref{fig:Nonmonotonous_MFPTs} are a direct consequence of this mutual water repulsion and the induced underdamped motion.

\section{Discussion}
\label{sec:Discussion}
\subsection{Fokker-Planck Kinetics with and without Diffusivity Profile}
\label{sec:Discussion_Diffusivities}
\begin{figure}
\includegraphics[width=.9\columnwidth]{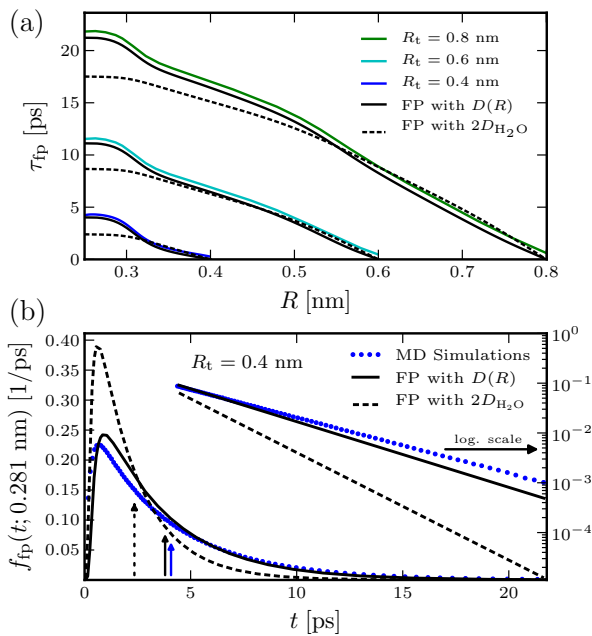}
\caption{\label{fig:mfpt_sim_fp}[Color online] 
(a)~MFPTs from MD simulations (\textit{solid color lines}, same data as in Fig.~\ref{fig:rel_diff}b), and from the FP description of 
Eq.~\ref{eq:MFPT} with constant diffusivity $2 D_\hho$ (\textit{dashed}) and with diffusivity profile $D(R)$ from Fig.~\ref{fig:rel_diff}c) (\textit{solid black lines}) for several target separations $R_\tx{t}$.
(b)~FPT distribution $f_\tx{fp}$ to reach a separation $R_t=0.4~\tx{nm}$  for water pairs starting within the first coordination shell at $R=0.281\pm0.001~\tx{nm}$. 
Histograms from MD simulations at $T=300~\tx{K}$ (\textit{blue dots}) are compared to the numerical solution of Eq.~\ref{eq:Fokker-Planck}  with constant diffusivity $2 D_\hho$ (\textit{dashed black line}) and with diffusivity profile $D(R)$ from Fig.~\ref{fig:rel_diff}c) (\textit{solid black line}). 
Data are shown on both linear and logarithmic vertical scales, vertical arrows indicate the mean of the corresponding distributions. FPT distributions for other target separations $R_\tx{t}$ and numerical details are found in App.~\ref{sec:FP_numerical_solution}.
}
\end{figure}

To what extent is this local friction increase relevant for the water-bond breakage kinetics? To quantify the relevance of the change in local friction, we compare in Fig.~\ref{fig:mfpt_sim_fp}a) MFPT curves from MD data already shown  in Fig.~\ref{fig:rel_diff}b) (\textit{colored lines}) with analytical predictions resulting from Eq.~\ref{eq:MFPT} using the diffusivity profiles $D(R)$ (\textit{solid lines}) shown in Fig.~\ref{fig:rel_diff}c) as well as calculations employing a constant diffusivity  $D=2D_\hho$ (\textit{broken lines}).
The solid lines by construction match the MD data nicely, the vertical shift being caused by the $20~\tx{fs}$ time discretization of the underlying MD trajectory, while the analytical predictions are calculated in the continuum (cf. App.~\ref{sec:MFPTs_time_resolution_model}). It is seen that the assumption of a constant
diffusivity leads to a considerable underestimate of the MFPTs. The time to reach the target separation $R_t=0.4~\tx{nm}$  from the first coordination shell ($R\lesssim0.28~\tx{nm}$) is underestimated
by a factor of roughly one half. 

The accuracy of the FP approach involving the diffusivity profile is demonstrated when comparing first-passage time (FPT) distributions:
Fig.~\ref{fig:mfpt_sim_fp}b) contrasts the FPT histogram from MD data for $R_\tx{t}=0.4~\tx{nm}$ with FPT distributions from the 
numerical solution of the FP equation (numerical details are given in App.~\ref{sec:FP_numerical_solution}), 
again using the flat diffusivity $2D_\hho$ and the actual diffusivity profile $D(R)$.
Only the FP approach including the $D(R)$ profile  correctly reproduces the entire  FPT distribution from MD simulations
and in particular also the exponential tail of the distribution, as shown by the plot using the  logarithmic scale on the right.
Systematic  discrepancies are observed on short  time scales $\lesssim1~\tx{ps}$ 
where  the MD data show more  ``fast'' transitions than the FP description.
These effects are caused by ballistic motion of water molecules and cannot be captured by a Markovian description.
FPT distributions corresponding to other target separations are shown in Fig.~\ref{fig:fptdist_sim_FP} in App.~\ref{sec:FP_numerical_solution}.

\subsection{Testing the Quality of the Reaction Coordinate}
\label{sec:RC-Test}
\begin{figure}
\includegraphics[width=.9\columnwidth]{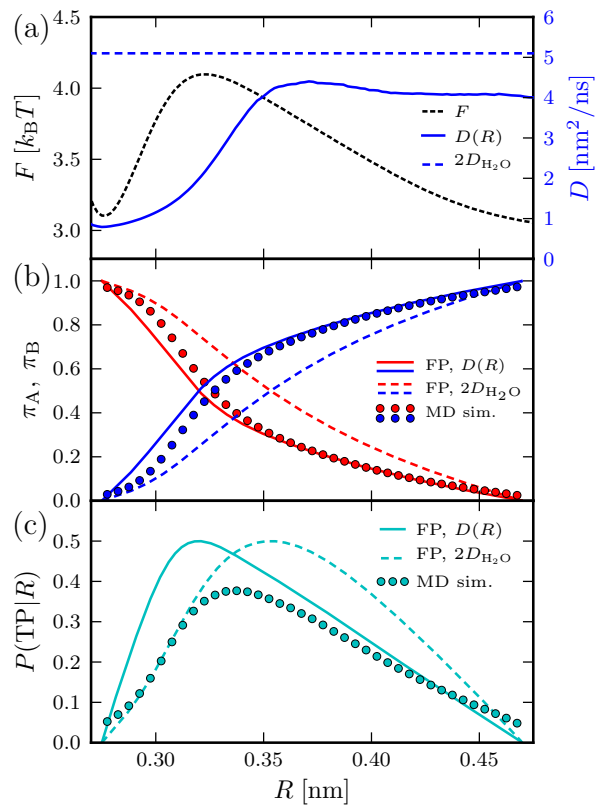}
\caption{\label{fig:react}[Color online] 
(a)~Free energy and diffusivity profiles in the range of separations between the first and second coordination shell for $T=300~\tx{K}$. 
(b)~Commitment probabilities $\pi_A$ and $\pi_B$.
(c)~Transition path probability $P(\tx{TP}\vert R)$.
Simulation data (\textit{circles}) are compared to 
FP estimates  based on the  diffusivity profile $D(R)$  (\textit{solid lines})
and  a constant  diffusivity $2D_\hho$  (\textit{broken lines}).
}
\end{figure}

Although our procedure does not strictly depend on the fact that the separation $R$ is a ''good'' RC,
the whole mapping on a one-dimensional FP equation is certainly more meaningful if it is. 
To check the quality of our RC, we divide $R$ into a bound region A for $R<R_\tx{A} = 0.275~\tx{nm}$,  
an unbound region B for  $R\geq R_\tx{B}=0.47~\tx{nm}$ and the intermediate region for $0.275~\tx{nm}\leq R< 0.47~\tx{nm}$ which roughly encompasses the free-energy barrier (see Fig.~\ref{fig:react}a)).
For a diffusive process described by the FP equation (Eq.~\ref{eq:Fokker-Planck}), the committor $\pi_\tx{X}(R)$ specifying the probability of first reaching region $X\in\{\tx{A},\tx{B}\}$ when starting from $R$ is a solution of the stationary backward FP equation~\cite{Gardiner}
\begin{equation}
 \label{eq:Backward_FP}
 \tx{e}^{\beta F(R)}\pdiff{R}\left(\tx{e}^{-\beta F(R)}D(R)\pdiffarg{\pi_\tx{X}(R)}{R}\right)=0.
\end{equation}
The committor in addition fulfills boundary conditions $\pi_\tx{X}(R_\tx{X})=1$ and $\pi_\tx{X}(R_\tx{Y})=0$, with Y$\neq$X. The solutions are given by
\begin{equation}
 \label{eq:Committor_solution}
  \begin{split}
  \pi_\tx{A}(R)&=\frac{1}{\C N}\int_{R}^{R_\tx{B}}\dd R^\prime\;\frac{\tx{e}^{\beta F(R^\prime)}}{D(R^\prime)},\\
  \pi_\tx{B}(R)&=1-\pi_\tx{A}(R)=\frac{1}{\C N}\int_{R_\tx{A}}^{R}\dd R^\prime\;\frac{\tx{e}^{\beta F(R^\prime)}}{D(R^\prime)},
 \end{split}
\end{equation}
with the normalization factor
\begin{equation}
\label{eq:Committor_normalization}
\C N\equiv\int_{R_\tx{A}}^{R_\tx{B}}\dd R^\prime\;\frac{\tx{e}^{\beta F(R^\prime)}}{D(R^\prime)}.
\end{equation}
In Fig.~\ref{fig:react}b) we show the committors $\pi_A(R)$ and $\pi_B(R)$, which specify the probability of first reaching region A and B, respectively, from MD data (circles) and from the exact solution of the FP equation (Eqs.~\ref{eq:Committor_solution} and \ref{eq:Committor_normalization}) employing constant bulk (broken line) or true diffusivity profile (solid line). Agreement between solid lines and MD data is quite good, although deviations close to region A are discernible and might point to residual barrier-crossing events orthogonal to the coordinate $R$~\cite{Csajka1998}.

For a one-dimensional diffusive system, the probability $P(\tx{TP}|R)$ that a path passing through $R$ is a transition path (TP), i.e., a path directly connecting the regions $A$ and $B$, is given by~\cite{Best2005}
\begin{equation}
\label{eq:TP_probability}
P(\tx{TP}|R)
=2\pi_\tx{A}(R)\pi_\tx{B}(R)=2\pi_\tx{A}(R)(1-\pi_\tx{A}(R)), 
\end{equation}
where the factor $2$ takes into account that a TP can start in A and reach B or vice versa.
The probability $P(\tx{TP}|R)$ reaches its maximum value $0.5$ at the transition state denoted by $R^\ddagger$~\cite{Hummer2004}, where $\pi_\tx{A}(R^\ddagger)=\pi_\tx{B}(R^\ddagger)=0.5$.
A ``good'' RC is characterized by a maximum value of the TP probability near this one-dimensional diffusive limit of 0.5~\cite{Best2005}. In contrast, for ``poor'' RCs, which do not single out the transition states, this maximum is considerably lower; the reason is that for ``poor'' RCs excursions from and to A and excursions from and to B dominate all along the coordinate such that TPs are rare everywhere between A and B~\cite{Geissler1999,Bolhuis2002}.

Committor and TP probabilities are estimated by analyzing all simulation paths within the region $R\in[R_\tx{A},R_\tx{B}]=[0.275~\tx{nm},0.47~\tx{nm}]$ within a time window of $100~\tx{ps}$ (time resolution $\delta t=0.01~\tx{ps}$).
$P(\tx{TP}|R)$ from MD data in Fig.~\ref{fig:react}c) reaches a maximal value of $P(\tx{TP}|R^\ddagger) \approx0.38$, the position $R^\ddagger$  being slightly displaced from the FP prediction employing the $D(R)$-profile (solid line) and a constantant $D$ (broken line) by about $0.02~\tx{nm}$; though caution is recommended in interpreting the TP probability test~\cite{Peters2010}, we conclude that the separation $R$ is an acceptable RC unlike in the similar problem of ion unbinding~\cite{Geissler1999}.


\subsection{Interpretation of Diffusivity Profiles}
\label{sec:Interpretation_Diffusivities}
Based on these findings, it is possible to give a quite intuitive interpretation of the diffusivity profile.
Assuming that the relative dynamics of two water molecules can be described as a diffusive process along a single path $\V R$ in the full high-dimensional configuration space, the projection onto one single coordinate, in this case the oxygen-oxygen separation $R$, generally leads to considerable changes in free-energy and diffusivity.

For convenience we assume the path $\V R(s),\;s\in[0,l]$ of total contour length $l$ being arc-length parametrized, i.e., $\abs{\dd \V R(s)/\dd s}=1\;\forall s$, and the reactive flux tube bing quite narrow such that the idea of a single, dominating path makes sense~\cite{E2010}.
The vector $\V R\equiv \left(R,\V R_\perp\right)$ is split up into the coordinate $R$ and an ortogonal, vectorial component $\V R_\perp$, implying
\begin{equation}
\label{eq:Path_phi}
\left|\frac{\dd \V R}{\dd s}\right|=\sqrt{\left|\frac{\dd \V R_\perp}{\dd s}\right|^2+\left(\frac{\dd R}{\dd s}\right)^2}=1\quad\forall s.
\end{equation}
We assume a one-to-one correspondence between the arc-length variable $s$ and the relative separation $R$, i.e., a path which does not take any value of $R$ more than once; in this case, the coordinate $R$ is just a reparametrization of $s$.
As is well-known~\cite{Weiss1966,Hinczewski2010JCP}, such a reparametrization can sensibly alter free-energy and diffusivity; more precisely the corresponding profiles along the coordinates $R$ and $s$ are connected via
\begin{equation}
\label{eq:Reparametrization}
\begin{split}
F(R)&=\tilde F(s)+\kBT\log{\left(\frac{\dd R}{\dd s}\right)},\\
D(R)&=\tilde D(s)\left(\frac{\dd R}{\dd s}\right)^2.
\end{split}
\end{equation}
Combining Eqs.~\ref{eq:Path_phi} and \ref{eq:Reparametrization} one deduces
\begin{equation}
\label{eq:dphi_perp_ds}
\left|\frac{\dd \V R_\perp}{\dd s}\right|=\sqrt{1-\left(\frac{\dd R}{\dd s}\right)^2}=\sqrt{1-\frac{D(R(s))}{\tilde D(s)}},
\end{equation}
meaning that the knowledge of the diffusivity profile $D(R)$ along the coordinate $R$ allows to draw conclusions on the shape of the path $\V R(s)$;
that is, a reduction of the diffusivity $D(R)$ along a chosen RC $R$ is a signature of pronounced contributions to the reaction path that are orthogonal to the RC.
Deviations of the diffusivity from the value $\tilde D(s)$ thus indicate a non-negligible component of the path perpendicular to $R$.
 Since only the magnitude of this perpendicular component can be accessed, while the direction remains uncertain, the path $\V R$ can not be completely reconstructed from the knowledge of $D(R)$. Defining
 \begin{equation}
\label{eq:phi_perp}
\begin{split}
R_\perp\equiv
\int_{R_0}^R\dd R^\prime\;\left|\frac{\dd \V R_\perp}{\dd R^\prime}\right|&=
\int_{R_0}^R\dd R^\prime\;\sqrt{\frac{\tilde{D}(s)}{D(R^\prime)}-1},
\end{split}
\end{equation}
ans using the relations in Eqs.~\ref{eq:Reparametrization} and \ref{eq:dphi_perp_ds}
however allows to visualize the path in the $(R,R_\perp)$-plane.
Choosing a constant $\tilde{D}(s) = 2 D_\hho$ for illustrative purposes, we show in Fig.~\ref{fig:TimeSeries}c) a fictitious path 
 in the plane $(R, R_\perp)$ that would be consistent with the diffusivity profile $D(R)$ actually extracted from MD simulations.  We observe that the pictorial reaction path has large contribution orthogonal to $R$ within 
 the first coordination shell, where the diffusivity profile shows its prominent drop, i.e., for relative separations $0.26~\tx{nm}\lesssim R\lesssim0.34~\tx{nm}$. Previous simulation results suggest that the orthogonal degree of freedom involved in water-bond breakage is in fact of angular nature~\cite{Svishchev1993,Laage2008}. 

\section{Conclusions}
\label{sec:Conclusions}
Summarizing, we have resolved the diffusivity profile for relative water dynamics based on a MFPT analysis of the stochastic trajectories obtained from MD simulations of SPC/E water. Over a wide range of temperatures, the diffusivity within the first coordination shell drops by a factor of more than six compared to large separations, where both partners diffuse independently. The form of the diffusivity profile is necessary to reproduce dynamic properties such as FPT distributions observed in the simulations and can be interpreted in terms of a reaction path which is distorted with repect to the resolved separation coordinate $R$.

We cautiously remark that a distorted reaction path is only one of a few mechanisms that would modify the local diffusivity;  orthogonal energetic barriers (which are, based on our results shown in Fig.~\ref{fig:react}, presumably small in the present case but dominate in related problems~\cite{Geissler1999}) and competing reaction paths~\cite{Noe2009} or flux-tube width variations~\cite{E2010,Berezhkovskii2011JCPa} are additional complications.
An understanding of the precise mechanisms at work when water molecules separate from each other is desirable, but requires the applications of more complex concepts, which we have not pursued in this paper; among others, transition path theory~\cite{E2010} or Markov-state modeling in full configuration space~\cite{Prinz2011} may prove useful.
The charm of our approach is that it allows for a  consistent description of the kinetics of water-bond breakage even without a detailed knowledge of the transition path and the involved degrees of freedom.

\begin{acknowledgments}
Financial support from the DFG (SFB 863) and from the Elitenetzwerk Bayern in the framework of CompInt (YvH, FS) is acknowledged.
\end{acknowledgments}

\appendix
\renewcommand{\thefigure}{\Alph{section}\arabic{figure}}

\section{Single Water Molecule Diffusion Coefficients}
\label{sec:H2O_Diffusion_Coefficients}
\setcounter{figure}{0}

\begin{figure}
\includegraphics[width=.9\columnwidth]{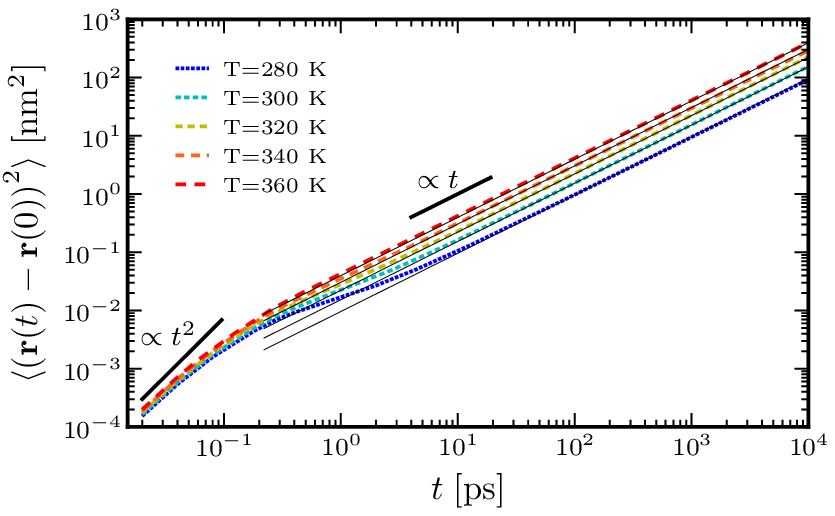}
\caption{\label{fig:MSD_single}[Color online] Single SPC/E water molecule MSDs for various temperatures (\textit{dashed color lines}) and best linear fits to the data (\textit{solid black lines}).}
\end{figure}

\begin{figure}
\includegraphics[width=.9\columnwidth]{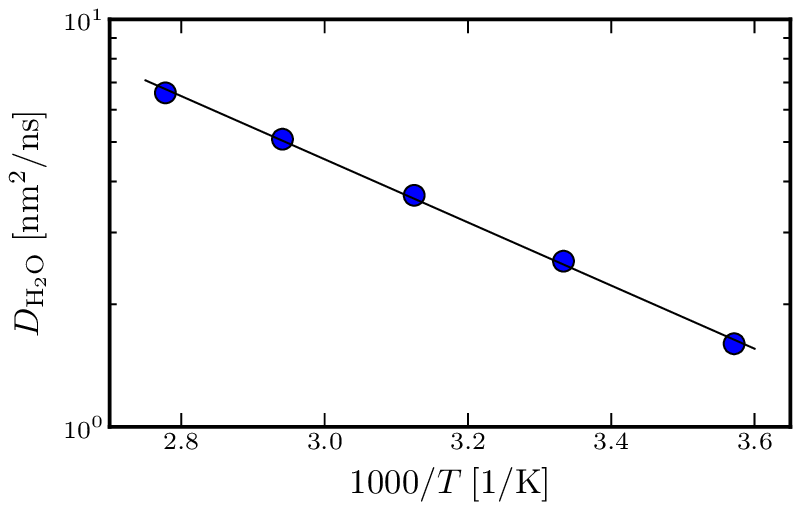}
\caption{\label{fig:Diff_single}[Color online] Arrhenius plot of the temperature dependence of the SPC/E diffusion coefficient: \textit{Symbols} denote results obtained through fits (see text) to the MSD data shown in Fig.~\ref{fig:MSD_single}, the \textit{line} shows that within the studied range of temperatures this dependence is well approximated by $D_\hho(T)\approx956\exp{\left(-1783.6~\tx{K}/T\right)}~\tx{nm}^2/\tx{ns}$.}
\end{figure}

Fig.~\ref{fig:MSD_single} shows the MSD of single SPC/E molecules extracted from MD simulations at various temperatures;
while the MSDs show a quadratic dependence on time characteristic for ballistic motion on the femtosecond time scale, a smooth crossover to a diffusive scaling is observed for $t\approx\C O(\tx{ps})$. The diffusion constant of a single water molecule
\begin{equation}
D_\hho=\lim_{t\to\infty}\frac{\Av{(\V r(t)-\V r(0))^2}}{6t}, 
\end{equation}
is obtained through linear fits to the data in the time range $10~\tx{ps}<t<10^3~\tx{ps}$.
In Tab.~\ref{tab:diff-constants} our results for the SPC/E diffusion coefficients are compared to results from experiments and other simulation studies.
As can be seen in Fig.~\ref{fig:Diff_single}, the diffusion coefficient shows an Arrhenius-like temperature dependence within the investigated range of temperatures in agreement with experimental findings~\cite{Price1999}.

\section{Mean First-Passage Times from Trajectories with Finite Time Resolution}
\label{sec:MFPTs_time_resolution_model}
\setcounter{figure}{0}
In Fig.~\ref{fig:time_resolution} we show several MFPT curves, which were obtained from the same simulation run by only varying the time resolution $\delta t$ employed for the MFPT analysis described in Sec.~\ref{sec:MFPTs_from_MD}.
\begin{figure}
\includegraphics[width=.9\columnwidth]{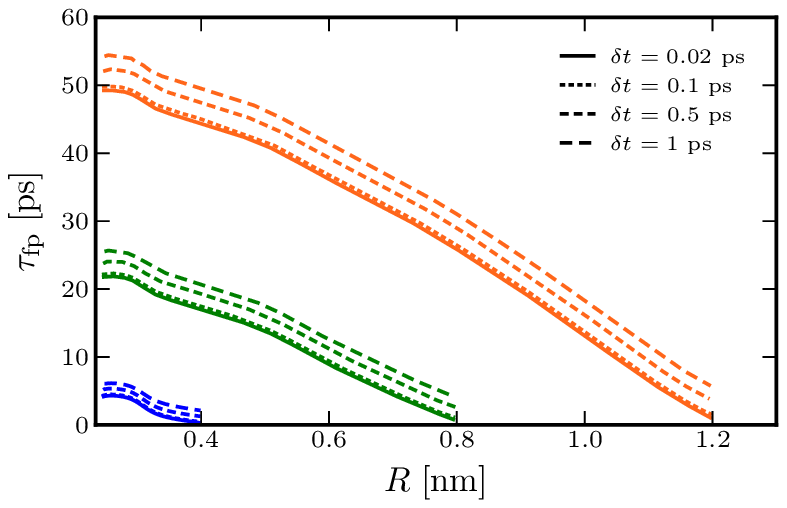}
\caption{\label{fig:time_resolution}[Color online] Dependence of the MFPT curves $\tau_\tx{fp}$ on the time resolution $\delta t$ of the underlying trajectories. Results for several target separations $R_\tx{t}=1.2~\tx{nm}$ (\textit{orange}), $R_\tx{t}=0.8~\tx{nm}$ (\textit{green}) and $R_\tx{t}=0.4~\tx{nm}$ (\textit{blue}) determined from MD simulation data at $T=300~\tx{K}$ are shown.
}
\end{figure}
The reason for the differences between the MFPT curves, are excursions beyond $R_\tx{t}$ and back, which are not registered due to the finite time resolution $\delta t$ and thereby affect the estimate of the mean.
Note that the curves in Fig.~\ref{fig:time_resolution} are mainly shifted vertically with respect to each other; we found that within the statistical uncertainty the choice of a specific time resolution had no visible effect on the form of the resulting diffusivity profile, which according to Eq.~\ref{eq:MFPTmethod} only depend on the slope of the MFPT curves (comparison not shown).

Note that the vertical shifts in Fig.~\ref{fig:time_resolution} are in fact much larger than the resolution $\delta t$ of the trajectories.
To demonstrate the influence of $\delta t$ on the observed MFPT curves, we use a simple model system: free diffusion of a particle along the coordinate $x$ with diffusion constant $D_\tx{p}$. A reflecting boundary at $x=0$ restricts the particle position to the positive part of the coordinate axis; the corresponding Green's function is given by
\begin{equation}
 G(x|x_0;t)=\frac{1}{\sqrt{4\pi D_\tx{p} t}}\left(\tx{e}^{-\frac{(x-x_0)^2}{4D_\tx{p}t}}+\tx{e}^{-\frac{(x+x_0)^2}{4D_\tx{p}t}}\right)\Theta(x).
\end{equation}
Since the process is Markovian, the probability distribution of finding the particle at position $x$ at time $t+\delta t$ is related to the probability distribution at time $t$ by
\begin{equation}
 \label{eq:Prob_deltat_later}
 P(x;t+\delta t)=\int_0^\infty\dd x^\prime\,G(x|x^\prime;\delta t)P(x^\prime;t).
\end{equation}
According to Eq.~\ref{eq:MFPT} the MFPT of reaching $x_\tx{t}$ when starting out from $x$ for a flat diffusivity $D(x)=D_\tx{p}$ and free-energy profiles $F(x)=\tx{const.}$ is given by
\begin{equation}
\label{eq:MFPT_Example_exact}
 \tau_\tx{fp}(x,x_\tx{t})=\tau_\tx{d}\left(1-\left(\frac{x}{x_\tx{t}}\right)^2\right),
\end{equation}
where the characteristic time $\tau_\tx{d}\equiv x_\tx{t}^2/(2D_\tx{p})$ for diffusion over the length $x_\tx{t}$ was defined.

\begin{figure}
\includegraphics[width=.9\columnwidth]{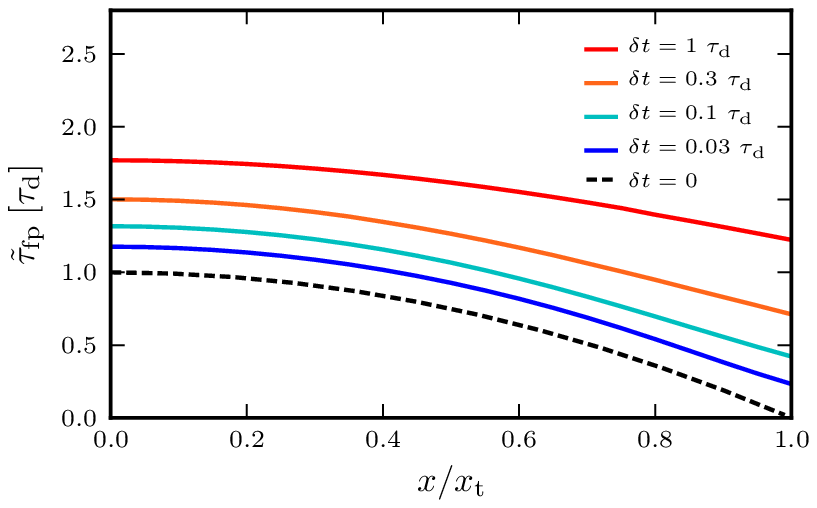}
\caption{\label{fig:MFPTs_time_resolution_model}[Color online] MFPT curves for reaching $x_\tx{t}$ in the case of free diffusion next to a reflecting wall at $x=0$ for different time resolution $\delta t$ of the underlying trajectory (see text for details). \textit{Solid colored lines} are obtained by the numeric procedure described in the text in App.~\ref{sec:MFPTs_time_resolution_model}, the \textit{dashed black line} denotes continuum MFPTs~(Eq.~\ref{eq:MFPT_Example_exact}). Times are given in units of the characteristic diffusion time $\tau_\tx{d}\equiv x_\tx{t}^2/(2D_\tx{p})$.
}
\end{figure}

We show in Fig.~\ref{fig:MFPTs_time_resolution_model} that MFPTs obtained from a trajectory with finite time resolution differ from the MFPT calculated in the continuum, because paths will eventually cross the target and come back many times before a position $x\geq x_\tx{t}$ is first recorded in the trajectory with finite time resolution. A lower bound for MFPT estimates based on trajectories with finite time resolution is obtained by the following numeric procedure:
\begin{enumerate}
\item At time $t=0$ the particle is located at $x_0$, thus the probability distribution reads $P(x;t=0)=\delta(x-x_0)$. Consequently, the probability of finding the particle left of the target is $P_\tx{left}=1$. Since no transitions beyond the target have been observed yet, the current MFPT estimate is $\tilde \tau_\tx{fp}=0$. The probability distribution at time $t=\delta t$ according to Eq.~\ref{eq:Prob_deltat_later} thus simply is $P(x)=G(x|x_0;\delta t)$ which is evaluated along $x$ with a resolution $\delta x=0.03~x_\tx{t}$.

The following steps are repeated until the exit condition is met:
\item%
Linearly interpolate the discrete values of $P(x)$ to obtain a continous function $\tilde P(x)$. The probability of still finding the particle left of the target is determined numerically by integration
\begin{equation}
P_\tx{left}^\tx{new}=\int_0^{x_\tx{t}}\dd x^\prime\,\tilde P(x).
\end{equation}
The fraction $\chi=P_\tx{left}-P_\tx{left}^\tx{new}$ of particles is thus found on the right of the target ($x>x_\tx{t}$) for the first time.
\item The fraction $\chi$ must have crossed the target between time $t-\delta t$ and time $t$ and contributes to the observed MFPT, which is updated accordingly:  $\tilde\tau_\tx{fp}=\tilde\tau_\tx{fp}+\chi(t-\delta t)$.
\item \textit{If} $P_\tx{left}^\tx{new}<0.001$, i.e., if less than 0.1~\% of the particles have not been observed right of the target at least once, \textit{then} exit the loop and return the MFPT estimate $\tilde\tau_\tx{fp}$. \textit{Else}:
\begin{enumerate}
 \item Numerically calculate the probability distribution at time $t+\delta t$ using Eq.~\ref{eq:Prob_deltat_later} on a discrete grid with $\delta x=0.03~x_\tx{t}$; herefore, start off with the interpolated version $P(x;t)=\tilde P(x)$ and choose $x_\tx{t}$ as upper integration limit in Eq.~\ref{eq:Prob_deltat_later} thereby neglecting the fraction of particles which reached separations $x\geq x_\tx{t}$ in the last iteration.
 \item Set $P_\tx{left}= P_\tx{left}^\tx{new}$ and $t=t+\delta t$.
 \item Go back to (2.).
\end{enumerate}
\end{enumerate}
MFPT curves resulting from the procedure described above are shown in Fig.~\ref{fig:MFPTs_time_resolution_model}: they show the same characteristics as the MFPT curves from MD simulation data in Fig.~\ref{fig:time_resolution}, namely increasing $\delta t$ shifts up vertically the curves; distortions of the curves are only observed for $\delta t\gtrsim \tau_\tx{fp}$.
As is clearly seen, one has $\tilde\tau_\tx{fp}>\tau_\tx{fp}+\delta t$ for the smaller values of $\delta t$: the deviations are thus \textit{not} caused because the first passage time is recorded with an uncertainty on the order of the time resolution, but because the first observed passage time in the discretely sampled trajectory can exceed by far the FPT in the continous trajectory.


\section{Numerical Solution of the Fokker-Planck Equation}
\label{sec:FP_numerical_solution}
\setcounter{figure}{0}
When discretizing the spatial coordinate $R$ into $N$ bins of width $\Delta R$, the FP equation (Eq.~\ref{eq:Fokker-Planck}) takes the form of a master equation~\cite{Bicout1998}
\begin{equation}
\label{eq:MasterEq}
\begin{split}
 \frac{\partial P_i(t)}{\partial t}&=W_{i,i-1} P_{i-1}(t)+W_{i,i+1} P_{i+1}(t)+W_{i,i} P_i(t),\\
 W_{i,i}&\equiv-W_{i-1,i}-W_{i+1,i}.
\end{split}
\end{equation}
Bin indices are denoted by $i\in\{1,2,\dots,N\}$, and the probability $P_i(t)$ of observing a relative separation $R$ within bin $i$, i.e., a separation $R\in [R_i-\Delta R/2,R_i+\Delta R/2)$, at time $t$ as well as the transition rates $W_{i,j}$ from bin $j$ to bin $i$ depend on both free-energy $F$ and diffusivity $D$
\begin{equation}
\label{eq:FP_Discretization}
W_{i,i+1}=\frac{D_i+D_{i+1}}{2\left(\Delta R\right)^2}\,\exp{\left(-\frac{F_i-F_{i+1}}{2\kBT}\right)},
\end{equation}
where $F_i\equiv F(R_i)$ and $D_i\equiv D(R_i)$. Due to detailed balance the transition rates between neighboring bins are not independent
\begin{equation}
 \label{eq:W}
 W_{i+1,i}=\exp{\left(-\frac{F_{i+1}-F_i}{\kBT}\right)}\, W_{i,i+1}.
\end{equation}
The linear transformation $\tilde P_i(t)=\exp{(\beta F_i/2)}P_i(t)$ with $\beta\equiv(\kBT)^{-1}$ converts Eq.~\ref{eq:MasterEq} into a simple differential equation involving a tridiagonal, symmetric matrix with entries $\tilde{W}_{ij}$
\begin{equation}
\label{eq:MasterEqSymmetric}
\frac{\partial\tilde{P}_i(t)}{\partial t}=\sum_{j=1}^N \tilde {W}_{ij} \tilde{P}_j(t)\quad\tx{with}\quad\tilde{W}_{ij}\equiv\tx{e}^{\beta F_i/2}\,W_{ij}\,\tx{e}^{-\beta F_j/2},
\end{equation}
which is readily solved in terms of a matrix exponential
\begin{equation}
\label{eq:MasterEqSymmetricSol}
\begin{split}
\tilde{P}_i(t)&=\sum_{j=1}^N\left(\tx{e}^{\tilde{\mathbf W} t}\right)_{ij}\,\tilde{P}_j(0)\quad \Leftrightarrow\\
P_i(t)&=\sum_{j=1}^N \tx{e}^{-\beta F_i/2} \left(\tx{e}^{\tilde{\mathbf W} t}\right)_{ij}\tx{e}^{\beta F_j/2}  P_j(0)\\
&\equiv \sum_{j=1}^{N}G_{ij}(t)P_j(0), 
\end{split}
\end{equation}
where in the last equality the Green's function $G_{ij}$ specifying the probability of landing in bin $i$ a time $t$ after the given start in bin $j$ was defined. The matrix exponential in Eq.~\ref{eq:MasterEqSymmetricSol} is computed numerically by diagonalization of the symmetric matrix $\tilde{\mathbf W}=\mathbf Q\mathbf\Lambda \mathbf Q^{-1}$, with $\mathbf Q$ being the matrix of eigenvectors and $\mathbf\Lambda$ being the diagonal matrix of eigenvalues of $\tilde{\mathbf W}$. The matrix exponential is then simply given by
\begin{equation}
\tx{e}^{\tilde{\mathbf W} t}=\mathbf Q\,\tx{e}^{\mathbf \Lambda t}\,\mathbf Q^{-1},\qquad\left(\tx{e}^{\mathbf \Lambda t}\right)_{ij}=\delta_{ij}\tx{e}^{\lambda_i t}.
\end{equation}
For the case of relative SPC/E water dynamics, a bin width $\Delta R=0.002~\tx{nm}$ was used; a reflective boundary condition ($W_{0,1}=W_{1,0}=0$) was imposed at $R_\tx{min}=0.235~\tx{nm}$ corresponding to a value of the free-energy $F\approx18~\kBT$.

FPT distributions are obtained by imposing an absorbing boundary condition at the target position $R_\tx{t}$, thus disregarding paths in the time evolution of $G_{ij}$ which have already reached the target beforehand: the total number $N$ of bins is chosen such that the target $R_\tx{t}$ is part of bin $N+1$ and the absorbing boundary condition is implemented by setting
\begin{equation}
 W_{N,N}=-W_{N-1,N}-W_{N+1,N},
\end{equation}
but neglecting the backward flux $W_{N,N+1}P_{N+1}(t)$ since $P_{N+1}(t)=0\;\forall t$. The survival probability $P_\tx{surv}^j$ for a start in bin $j$ is
\begin{equation}
 P_\tx{surv}^j(t)=\sum_{i=1}^{N}G_{ij}(t)=\sum_{i=1}^{N}\tx{e}^{-\beta F_i/2}\left(\tx{e}^{\tilde{\mathbf W}t}\right)_{ij}\tx{e}^{\beta F_j/2},
\end{equation}
which is evaluated numerically at times $t\in[0,200~\tx{ps}]$ with time resolution $\delta t=0.1~\tx{ps}$. The first-passage time (FPT) distribution is approximated by the finite difference
\begin{equation}
\label{eq:FPT_Dist_FP}
  f_\tx{fp}(t+\delta t/2;j)\approx\frac{P_\tx{surv}^j(t)-P_\tx{surv}^j(t+\delta t)}{\delta t}.
\end{equation}
The MFPT is obtained by taking the first moment of the FPT distribution,
\begin{equation}
\begin{split}
\tau_\tx{fp}&=\int_{0}^{\infty}\dd t\;t f_\tx{fp}(t)=\int_{0}^\infty\dd t\; P_\tx{surv}(t)\\
&\approx\delta t\left(\frac{1+P_\tx{surv}^j(M\,\delta t)}{2}+\sum_{n=1}^{M-1} P_\tx{surv}^j(n\,\delta t)\right), 
\end{split}
\end{equation}
where in our case $M=2000$. 

\begin{figure*}
\includegraphics[width=.9\textwidth]{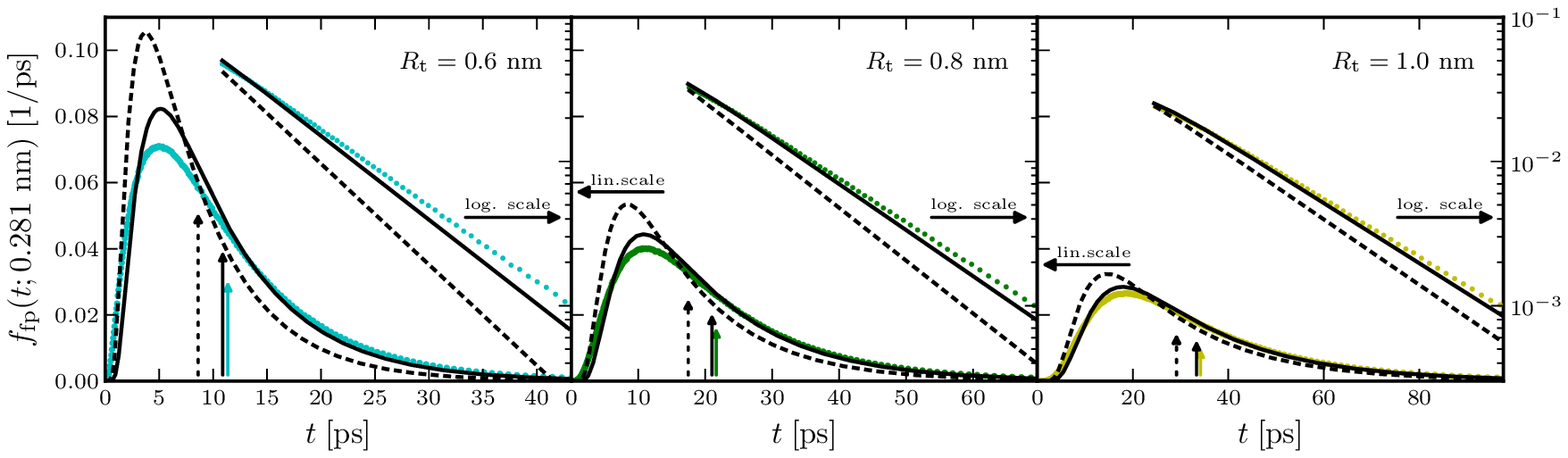}
\caption{\label{fig:fptdist_sim_FP}[Color online] First-passage time distributions $f_\tx{fp}$ for water pairs starting within the first coordination shell ($R_0=0.281~\tx{nm}$): Histograms from MD simulations at $T=300~\tx{K}$ (colored dots) are compared to the numerical solution of the FP equation (Eq.~\ref{eq:FPT_Dist_FP}) with flat diffusivity $2D_\hho$ (\textit{dashed black lines}) and with diffusivity profiles $D(R)$ from Fig.~\ref{fig:rel_diff}c) (\textit{solid black line}) for target separations $R_\tx{t}$ increasing from $0.6~\tx{nm}$ to $1.0~\tx{nm}$ from left to right. Data are shown on both linear and logarithmic vertical scales, vertical arrows indicate the mean of the corresponding distributions.
}
\end{figure*}

FPT histograms from MD data and from the numerical solution of the FP equation described in this section for other target separations than the one shown in Fig.~\ref{fig:mfpt_sim_fp}b) are found in Fig.~\ref{fig:fptdist_sim_FP}.
No significant impact on the FPT distributions was observed when refining the discretization in space and/or time.

%

\end{document}